# Epitaxial Growth of an *n*-type Ferromagnetic Semiconductor CdCr$_2$Se$_4$ on GaAs(001) and GaP(001)

Y.D. Park,[a,b)] A.T. Hanbicki,[a)] J. E. Mattson[c)] and B.T. Jonker[d)]
*Naval Research Laboratory*, Washington DC 20375

ABSTRACT

We report the epitaxial growth of CdCr$_2$Se$_4$, an *n*-type ferromagnetic semiconductor, on both GaAs and GaP(001) substrates, and describe the structural, magnetic and electronic properties. Magnetometry data confirm ferromagnetic order with a Curie temperature of 130 K, as in the bulk material. The magnetization exhibits hysteretic behavior with significant remanence, and an in-plane easy axis with a coercive field of ~125 Oe. Temperature dependent transport data show that the films are semiconducting in character and n-type as grown, with room temperature carrier concentrations of *n* ~ 1 x 10$^{18}$ cm$^{-3}$.

a) National Research Council Postdoctoral Fellow at NRL
b) current address: School of Physics and CSCMR, Seoul National University, Seoul 151-747, Korea.
c) George Washington University Postdoctoral Associate
d) Author to whom correspondence should be addressed; electronic mail: jonker@nrl.navy.mil

Ferromagnetic semiconductors (FMSs) provide unprecedented opportunity to tune and optimize spin-dependent behavior in semiconductor device heterostructures. Most efforts have focused on III-Mn-V materials such as GaMnAs,[1,2] which are *p*-type only since Mn acts as an acceptor in a III-V host. An *n*-type FMS material is of particular interest, since efficient electrical spin injection[3,4] and very long spin scattering lengths[5,6,7] have been confirmed for *electrons* rather than holes in semiconductors such as GaAs. In addition, modern high frequency and low power devices are based on electron transport. *N*-type behavior has been reported for new FMS materials such as $TiO_2$[8] and $CdMnGeP_2$.[9] However, these films were grown on $LaAlO_3$ / $SrTiO_3$ and $CdGeP_2$ substrates, respectively, and it is highly desirable to realize an *n*-type FMS material which can be epitaxially grown on a device-quality substrate compatible with existing electronics.

$CdCr_2Se_4$ is one of the chalcogenide FMS materials which were extensively studied several decades ago.[10,11] It is a spinel ferrite with the $AB_2X_4$ structure (56 atoms per unit cell), in which the Se anions form a cubic close-packed lattice, with the Cd cations occupying the resultant tetrahedral sites and the Cr cations occupying the octahedral sites. Slight lattice distortions and cation valence or site disorder are believed to play an important role in determining the properties of this class of materials. $CdCr_2Se_4$ has a lattice constant of 10.721 Å and a direct band gap of ~ 1.5 eV, and has been grown in bulk and polycrystalline thin film form. Until recently,[12,13] interest in these compounds languished because of the inability to incorporate them with established semiconductor device materials. We report here the epitaxial growth of *n*-type FMS $CdCr_2Se_4$(001) films on both GaAs and GaP(001) substrates, and describe the structural, magnetic and electronic properties.

The samples were grown by molecular beam epitaxy from elemental Knudsen cell style sources, with the surface structure monitored by reflection high energy electron diffraction (RHEED) during growth. Stoichiometric films were obtained for flux ratios Cd:Cr:Se of 1 : 4.6 : 36 (as measured by an ion gauge placed in the sample position) for substrate temperatures below 400˚C, with a typical growth rate of 5-6 Å/min. Cd was not incorporated at higher substrate temperatures, and the resulting film became polycrystalline with a composition consistent with $Cr_3Se_4$.[14]

Single crystal (001) films were obtained on both GaP and GaAs despite a lattice mismatch of 1.7% and 5.2% (tensile strain), respectively.[15] A 1x1 RHEED pattern appeared only when the fluxes were adjusted to provide stoichiometric $CdCr_2Se_4$ film growth. The quality of the pattern depended upon the character of the initial substrate surface. For growth on GaP(001) substrates where the surface oxide was simply thermally desorbed in ultra-high vacuum, the RHEED pattern disappeared within two hours of growth. When the GaP substrate was first treated with an $(NH_4)S$-based chemical passivation process,[16] the substrate surface order was improved (Figure 1a), and the RHEED pattern produced by the film retained a sharp but rather spotty character throughout 1000Å of growth, as shown in Figure 1b. Similar results were obtained for epitaxial growth on oxide desorbed GaAs(001) substrates. In this case, the RHEED pattern corresponding to $CdCr_2Se_4$(001) was again more diffuse and spotty than that of the initial substrate surface, but retained a clear single crystal character during the growth. The results above were largely independent of whether the $CdCr_2Se_4$ growth was initiated with a Cd or Se exposure of the substrate surface.

The initial film nucleation and crystalline order were significantly improved for growth on an MBE-grown GaAs buffer layer, which was grown on a GaAs(001) wafer in an attached III-V MBE chamber. In this case, the 2x4 pattern of the As-dimer terminated GaAs epilayer surface (Figure 1c) was immediately replaced by the 1x1 pattern of the $CdCr_2Se_4$ (001) film (Figure 1d), which persisted for the full duration of the growth.

The film structure, orientation and composition were confirmed by post-growth x-ray diffraction and fluorescence measurements. Theta-2theta scans show only peaks consistent with the $CdCr_2Se_4$ epilayer and the GaAs substrate. The position of the (004) $CdCr_2Se_4$ peak indicates an out-of-plane lattice parameter of 10.708 Å, which is 0.38% smaller than the reported bulk value (10.721 Å), and consistent with in-plane tensile strain expected from the lattice mismatch with GaAs.

The epitaxial films show magnetic and transport properties comparable to the bulk and largely independent of the substrate. The temperature dependence of the magnetization is shown in Fig. 2a as measured with a 1 kOe field applied along an in-plane <110> axis. Curie temperatures determined from fits to the data are 130-132 K, equal to the value reported for bulk material. Magnetization data as a function of applied field (figure 2b) indicate the easy axis to be in-plane with a coercive field of approximately 125 Oe and remanence that is one half of the saturation magnetization. The saturation magnetization was found to be approximately 300 emu/cm$^3$, which corresponds to 2.5 $\mu_B$ per Cr atom or 5 $\mu_B$ per formula cell, assuming 16 Cr atoms per unit cell and the lattice parameter found from the x-ray diffraction data. This value compares well with experimental values reported in the literature of 5.6 $\mu_B$ per formula cell, and a recent theoretical calculation of ~ 6.2 $\mu_B$ per formula cell.[17,18]

One of the most attractive properties of $CdCr_2Se_4$ as a FMS is the potential to dope it either *n*- or *p*-type.[10] In addition, FM order is stabilized even in the insulating phase, which corresponds to a stoichiometric material with minimal cation site disorder. This is in marked contrast to the III-Mn-V FMS materials which are *p*-type only – since Mn acts as both the magnetically active element and as an acceptor in the III-V host, *and* a high hole density is critical in stabilizing FM order, it is not possible to independently control the FM alloy concentration and carrier type or density. Room temperature Hall measurements show the nominally undoped $CdCr_2Se_4$ films grown on undoped GaAs and GaP substrates to be *n*-type with an electron concentration of ~ 1-3 x $10^{18}$ $cm^{-3}$ . The Hall measurements were performed in the van der Pauw geometry using both soldered and annealed indium contacts, bias currents of 0.1-100 µA and an applied field of ± 1500 Oe.

Further confirmation of *n*-type behavior is provided by the temperature dependence of the resistivity. For bulk *n*-type $CdCr_2Se_4$, the resistivity exhibits a local maximum near the magnetic ordering temperature, which has been interpreted in terms of spin disorder scattering, while data for *p*-type $CdCr_2Se_4$ show only a slight change in slope near $T_C$.[10] Resistivity measurements were performed using standard *ac* techniques with a Quantum Design PPMS 6000 and a 4 point probe geometry, where the temperature was swept at 0.5 K/min. Data obtained from an unintentionally doped $CdCr_2Se_4$ film 1000 Å thick grown on a S-passivated GaP(001) substrate are shown in Figure 3 and exhibit a pronounced local maximum near 112 K. The corresponding magnetization data are superposed for reference, and indicate $T_c$ = 130 K. It is not clear at this point whether the difference in resistivity maxima and corresponding $T_c$ is a manifestation of a thin film phenomenon or a consequence of the relatively high carrier concentration.

This work was supported by the Office of Naval Research and the DARPA *Spins in Semiconductors* Program. YDP and ATH gratefully acknowledge support by ONR during their tenure as National Research Council Postdoctoral Fellows at NRL.

**Figure Captions:**

**Figure 1**. RHEED patterns during the growth of $CdCr_2Se_4$(001): *a)* initial S-passivated GaP(001) surface, [110] azimuth; *b)* after 30 minutes or 150 Å of $CdCr_2Se_4$ film growth; *c)* initial epi-GaAs(001) surface showing the 4-fold reconstruction of the 2x4-As surface (the azimuth is slightly off $[1\bar{1}0]$); *d)* after 4 minutes or 20 Å of $CdCr_2Se_4$ film growth.

**Figure 2.** a) Magnetization *vs* temperature for $CdCr_2Se_4$ on S-passivated GaP(001) (open circles) and on an epitaxial film of GaAs(001) (crosses). b) Magnetic hysteresis loops for $CdCr_2Se_4$/GaAs (001) for applied field parallel and perpendicular to the film plane showing the magnetic easy-axis to be in-plane.

**Figure 3.** Temperature dependence of the resistivity for a 1000 Å thick film on GaP(001). A pronounced maximum is observed near 112 K, confirming *n*-type behavior. The temperature dependence of the magnetization (obtained from SQUID magnetometry measurements) is shown for reference.

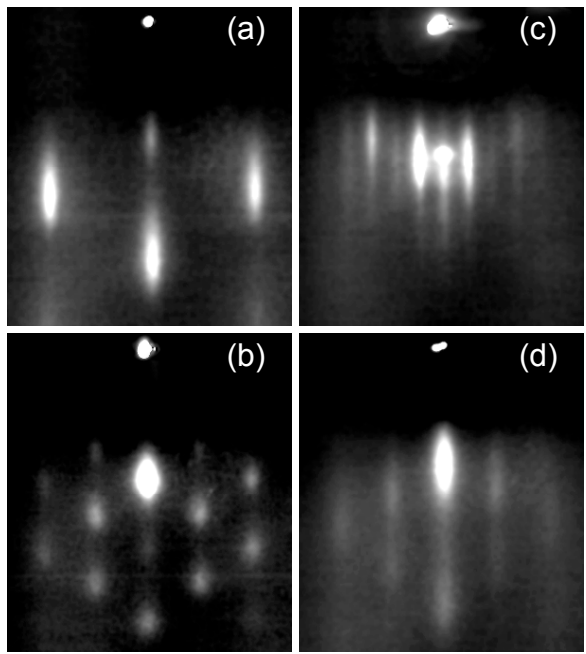

Figure 1 — Y.D. Park *et al.*

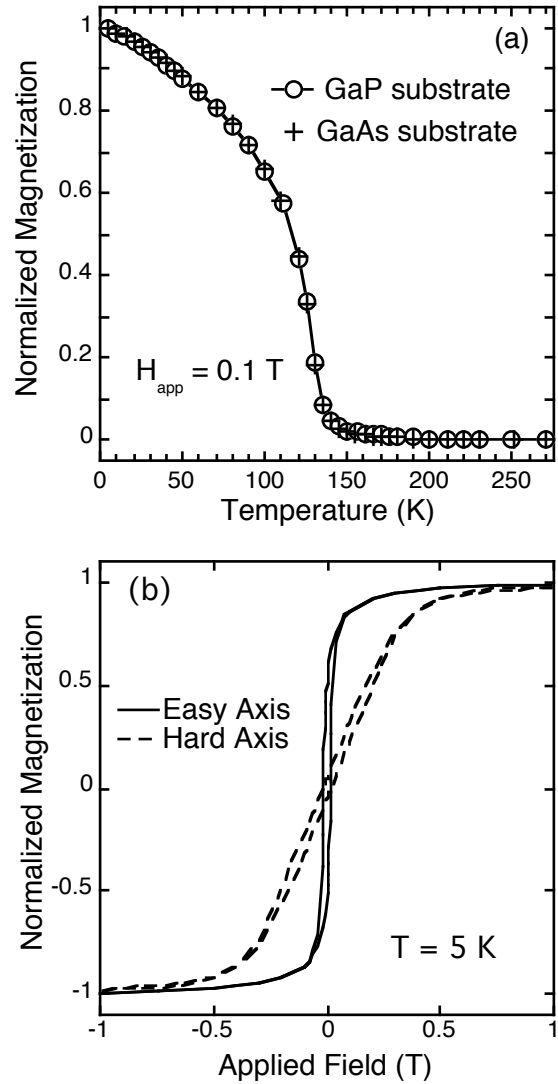

Figure 2 — Y.D. Park, *et al*.

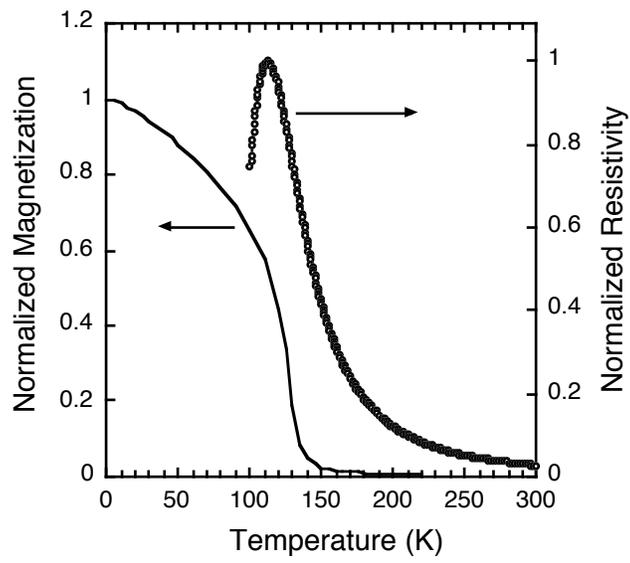

Figure 3 — Y.D. Park, *et al.*